\documentstyle[12pt]{article}

\renewenvironment{thebibliography}[1]
{\normalsize
 \begin{list}{[\arabic{enumi}]}
 {\usecounter{enumi} \setlength{\parsep}{0pt}
  \setlength{\itemsep}{3pt} \settowidth{\labelwidth}{[#1]}
  \sloppy}}
{\end{list}}

\renewcommand{\theequation}{\thesection.\arabic{equation}}
\newcommand{\cleqn}{\setcounter{equation}{0}}

\newcommand{\pr}{\hspace{\parindent}}
\newcommand{\bea}{\begin{eqnarray}}
\newcommand{\eea}{\end{eqnarray}}
\newcommand{\beq}{\begin{equation}}
\newcommand{\eeq}{\end{equation}}

\def\msbar{\ifmmode{\overline{\rm MS}} \else{$\overline{\rm MS}$} \fi}
\def\drbar{\ifmmode{\overline{\rm DR}} \else{$\overline{\rm DR}$} \fi}

\begin{document}

\hfill\vbox{\baselineskip14pt
            \hbox{KEK-TH-383}
            \hbox{KEK Preprint 93-182}
            \hbox{UT-665}
            \hbox{January 1994}
            \hbox{July 1994(revised)}}
\vspace{10mm}

\baselineskip22pt
\begin{center}
\Large	Two-loop renormalization group equations for
soft SUSY breaking scalar interactions: supergraph method
\footnote{Work supported in part by
Soryushi Shogakukai.}
\end{center}
\vspace{10mm}

\begin{center}
\large 	Youichi~Yamada
\end{center}
\vspace{0mm}

\begin{center}
\begin{tabular}{c}
{\it Theory Group, KEK, Tsukuba, Ibaraki 305, Japan}\\
\\
and\\
\\
{\it Department of Physics, University of Tokyo, Bunkyo-ku,
Tokyo 113, Japan}\\
\end{tabular}
\end{center}

\vspace{20mm}
\begin{center}
\large Abstract
\end{center}
\begin{center}
\begin{minipage}{13cm}
\baselineskip=22pt
\noindent
We obtain the one- and two-loop renormalization group equations
for soft SUSY breaking
scalar interactions in general, semi-simple SUSY gauge model,
by using the supergraph method.
We find that the method simplifies the calculation significantly
because of the non-renormalization theorem and also because
of the property that
the relevant divergences are derived by simple algebra
from those in exact SUSY case.
A disagreement with the existing result is found in
$\beta^{(2)}(m^2)$ and its cause is briefly discussed.
\end{minipage}
\end{center}
\vfill
\newpage

\baselineskip=22pt
\normalsize
\setcounter{footnote}{0}

\section{Introduction}
\pr
The gauge theory with softly broken supersymmetry (SUSY)
has been widely studied as one of the plausible
extension of the standard model.
The model contains many new couplings, the soft SUSY breaking interactions,
which are arbitrary in the low energy effective theory.
When the model is embedded in some unified theories,
such as the grand unified models and the supergravity models,
the soft SUSY breaking couplings are expressed in terms of a few
parameters at the very high unification scale. To obtain experimental
predictions, we should extrapolate the values of these couplings
to the weak scale by using the renormalization group equations.

The one-loop renormalization group equations for soft SUSY breaking
interactions have been given in \cite{1looplam} for general cases.
However, the two-loop contributions to the equations have not been known
for a long time, although they can in principle be derived from
the general $\beta$-functions in
refs.\cite{mv,2loopdr}. Very recently, the two-loop $\beta$-functions
for the gaugino masses \cite{2loopmg1,2loopmg2} and the scalar
interactions \cite{mv5} were obtained.
In this paper, we present an alternative method for calculating
the $\beta$-functions, the supergraph method, in case of
soft SUSY breaking scalar interactions.

The supergraph method \cite{supergraph1,supergraph},
which is a very powerful tool for studying exact SUSY models,
is also applicable to softly broken
SUSY models by using the `spurion' external
field \cite{soft,superfield2}.
This method has many nice features: manifest SUSY and divergence
cancellation, reduction of number of graphs and Lorentz indices,
and the non-renormalization theorem \cite{nonrenorm}.
In most previous works \cite{1looplam,2loopmg1,2loopmg2,mv5}, however,
the renormalization group equations for
soft SUSY breaking interactions have been calculated in the component
field method in the Wess-Zumino gauge.
Although ref.\cite{chang} has given a supergraph calculation
of the one-loop scalar interactions, unfortunately,
their specific method works only for very simple models.

We find that by using the supergraph method, the calculation of the
$\beta$-functions for the soft SUSY breaking scalar
interactions becomes much simpler than that in the component field method,
due to the non-renormalization theorem.
If the gauge group has no U(1) factor, we only need the divergent
parts of two- and one-point effective vertex functions of
chiral supermultiplets with the spurion insertions.
Moreover, these vertex functions can be obtained
by simple algebra from those in exact SUSY case,
at least to the two-loop order.

The paper is organized as follows. In section 2, we review
the superfield expression of the gauge model with softly broken SUSY.
A problem for the renormalization of
soft SUSY breaking terms is discussed.
In section 3, we show that the renormalization of soft SUSY breaking
scalar interactions can be obtained from the divergent parts of
two-point functions of chiral supermultiplets in exact SUSY case,
if the gauge group has no U(1) factor.
In section 4, we check that our method gives the correct
one-loop $\beta$-functions for
soft SUSY breaking scalar interactions.
In section 5, we calculate the two-loop $\beta$-functions for these
scalar interactions and compare them with the results in ref.\cite{mv5}
which are obtained by the component field method.
A discrepancy is found in the $\beta$-function for the scalar masses
and its possible cause is discussed.
Finally, section 6 gives our conclusions.


\section{Softly broken SUSY model in
the superfield formalism}
\pr
In this section, we review the superfield expression of the lagrangian
for general gauge model with softly broken SUSY. We then discuss
a problem in the renormalization of soft SUSY breaking terms
in the supergraph method and show that it is solved by a
suitable redefinition of superfields.

We first write down the lagrangian of general gauge model with
softly broken SUSY in the superfield formalism, by using
the spurion external field $\eta\equiv\theta^2$.
Our notation and convention for the superfield formalism are given
in the Appendix. We consider a
model with a semi-simple gauge group $G=\prod_A G_A$,
where $G_A$'s are simple subgroups.
In this paper, we assume that there is no U(1) factor in $G$,
which is crucial for later discussion.
The model contains chiral supermultiplets $\Phi_i$ in
the representations $R_i^A$ for the subgroup $G_A$ and
vector supermultiplets $V^A$.
Here we use the abbreviations: the index $A$ represents
the subgroups $G_A$ and their generators $T^A$, and $i$
represents the irreducible gauge multiplets and
their components.

The lagrangian is written as \cite{superfield1,soft}
\beq
{\cal L}={\cal L}_{SUSY}+{\cal L}_{soft}+{\cal L}_{GF}+{\cal L}_{FP},
\label{eq21}
\eeq
where the SUSY part is
\bea
{\cal L}_{SUSY}&=& \int d^4\theta \bar{\Phi}^i(e^{2gV})_i^j\Phi_j+
\frac{1}{4}\int d^2\theta W^{A\alpha}W^A_{\alpha}
+\frac{1}{4}\int d^2\bar{\theta}
\bar{W}^A_{\dot{\alpha}}\bar{W}^{A\dot{\alpha}} \nonumber \\
&&+\int d^2\theta (\frac{1}{6}\lambda^{ijk}\Phi_i\Phi_j\Phi_k+
\frac{1}{2}M^{ij}\Phi_i\Phi_j+L^i\Phi_i)+{\rm h.c.}, \label{eq22} \\
W_{\alpha}^A&=&\frac{1}{8g_A}\bar{D}^2\left[ e^{-2gV}D_{\alpha}e^{2gV}
\right] ^A, \nonumber \\
gV&=&g_AV^AT^A, \nonumber
\eea
and the soft SUSY breaking part is\footnote{Usual definitions
of $A$, $B$, $C$ are our $A^{ijk}/\lambda^{ijk}$, $B^{ij}/M^{ij}$,
$C^i/L^i$, respectively.}
\bea
{\cal L}_{soft}&=&-\int d^2\theta\eta
(\frac{1}{6}A^{ijk}\Phi_i\Phi_j\Phi_k+
\frac{1}{2}B^{ij}\Phi_i\Phi_j+C^i\Phi_i+\frac{m_A}{2}
W^{A\alpha}W^A_{\alpha})+{\rm h.c.}\nonumber \\
&&-\int d^4\theta \bar{\eta}\eta\bar{\Phi}^i(m^2)_i^j
(e^{2gV})_j^k\Phi_k.
\label{eq23}
\eea
An appropriate trace for group generators is understood in
all the lagrangians in this paper.
${\cal L}_{soft}$ contains the scalar masses $m^2$, the gaugino masses
$m_A$ and scalar interactions $A$, $B$, $C$.
The factor $e^{2gV}$ in (\ref{eq23}) is necessary to make
${\cal L}_{soft}$ invariant under the super-gauge transformation.
If $G$ has no U(1) factor, the Fayet-Iliopoulos term
$\int d^4\theta V$ is not generated.
The explicit forms of the gauge-fixing term
${\cal L}_{GF}$ and Fadeev-Popov ghost term
${\cal L}_{FP}$ are given in section 3.

We encounter one problem in the renormalization of (\ref{eq23}). The
lagrangian (\ref{eq23}) does not
contain all possible divergent terms involving $\eta$, $\bar{\eta}$.
In fact, loop correction produces the following types of
divergences \cite{soft,chang}, in addition to those in (\ref{eq23}),
\beq
\int d^4\theta[ \eta\bar{\Phi}\Phi,\;\; \bar{\eta}\Phi,\;\;
\bar{\eta}\eta\Phi,\;\; \bar{\eta}\eta D^{\alpha}W_{\alpha} ].
\label{eq24}
\eeq
The last term in (\ref{eq24}) can appear only if $G$ has a U(1) factor.

In the component field formalism, these terms can be expressed as linear
combinations of the usual terms in (\ref{eq23}) by eliminating
auxiliary components. But in the superfield formalism,
this procedure means the substitution of the equation of
motion into (\ref{eq24}) and is not easy to justify.
So, to obtain manifestly finite effective action in the
superfield formalism,
we should include the first three terms of (\ref{eq24})
to the original lagrangian :
\bea
{\cal L}&=&{\cal L}_{SUSY}+{\cal L}'_{soft}+{\cal L}_{GF}+{\cal L}_{FP},
\nonumber \\
{\cal L}'_{soft}&=&-\int d^2\theta\eta
(\frac{1}{6}\hat{A}^{ijk}\Phi_i\Phi_j\Phi_k+
\frac{1}{2}\hat{B}^{ij}\Phi_i\Phi_j+\hat{C}^i\Phi_i
+\frac{m_A}{2}W^{Aa}W^A_a)+{\rm h.c.}\nonumber \\
&&+\int d^4\theta[-\bar{\eta}\eta
\bar{\Phi}^i(\hat{m}^2)_i^j(e^{2gV})_j^k\Phi_k
+\bar{\Phi}^i(\eta\kappa_i^j+\bar{\eta}\kappa_i^{*j})
(e^{2gV})_j^k\Phi_k \nonumber \\
&&+\bar{\eta}\eta(\rho_1^i\Phi_i+\rho^*_{1i}\bar{\Phi}^i)
+(\bar{\eta}\rho_2^i\Phi_i+\eta\rho^*_{2i}\bar{\Phi}^i)].
\label{eq25}
\eea
Here we have introduced new couplings $\kappa$, $\rho_1$
and $\rho_2$, whose dimensions are 1, 3 and 2, respectively.

Fortunately, the lagrangian (\ref{eq25}) can be transformed into the
conventional form (\ref{eq23}) by the following
$\theta$-dependent field redefinition,
\bea
\Phi_i&=&\Phi'_i-\eta(\kappa_i^j\Phi'_j+\rho^*_{2i}),
\label{eq26}
\eea
Then the relations between the coupling constants in ${\cal L}_{soft}$ and
those in ${\cal L}'_{soft}$ are as follows:
\bea
A^{ijk}&=&\hat{A}^{ijk}+\lambda^{ljk}\kappa_l^i
+\lambda^{ilk}\kappa_l^j+\lambda^{ijl}\kappa_l^k, \label{eq28} \\
B^{ij}&=&\hat{B}^{ij}+M^{lj}\kappa_l^i+M^{il}\kappa_l^j
+\lambda^{ijl}\rho_{2l}^*, \label{eq29} \\
C^i&=&\hat{C}^i+L^{j}\kappa_j^i+M^{il}\rho_{2l}^*
+\rho_2^l\kappa_l^i-\rho_1^i, \label{eq210} \\
(m^2)_i^j&=&(\hat{m}^2)_i^j+\kappa^{*k}_i\kappa_k^j. \label{eq211}
\eea
In (\ref{eq210}), we have used the following identity
\beq
\int d^4\theta\bar{\eta}\eta\Phi=\int d^2\theta\eta\Phi. \label{eq212}
\eeq
Obviously, other coupling constants in (\ref{eq25}) are not affected
by the redefinition (\ref{eq26}).

The renormalization of the interactions (\ref{eq23}) can now be done
as follows.
First, we renormalize the lagrangian (\ref{eq25}) under two conditions:
that all the couplings in (\ref{eq25}) are independent of each other
and that the renormalized $\kappa$, $\rho_1$ and $\rho_2$ are set to 0.
Then the bare couplings ($\hat{A}$, $\hat{B}$, $\hat{C}$, $\hat{m}^2$,
$\kappa$, $\rho_1$, $\rho_2,\ldots$) in (\ref{eq25}) are expressed
in terms of the renormalized couplings ($A$, $B$, $C$, $m^2,\ldots$)
in (\ref{eq23}). Second, we obtain the bare couplings ($A$, $B$, $C$,
$m^2,\ldots$) in (\ref{eq23})
by using the relations (\ref{eq28}--\ref{eq211}).

If the gauge group has a U(1) factor, we further need an additional term
\beq
\rho_3^A\int d^4\theta\bar{\eta}\eta D^{\alpha}W_{\alpha}^A=
\frac{\rho_3^A}{4}\int d^4\theta(\bar{D}^2\bar{\eta})
(D^2\eta)V^A, \label{eq213}
\eeq
in the lagrangian (\ref{eq25}). By the field redefinition
$V_A=V'_A-\bar{\eta}\eta\rho_3^A$, we can absorb (\ref{eq213})
into $m^2$ as
\beq
\Delta(m^2)_i^j=2g_A(T^A)^j_i\rho_3^A. \label{eq214}
\eeq
The appearance of the $\rho_3$ term complicates the following discussions,
and we will discuss its consequences elsewhere.


\section{Evaluation of divergent parts}
\cleqn
\pr
In this section, we consider the calculation of the divergent
supergraphs which are
relevant to the renormalization of soft SUSY breaking scalar
interactions. We show that to the two-loop order,
the relevant divergences involving spurion $\eta$
are derived by simple algebra from those in exact SUSY case.
We follow the method given in ref.\cite{supergraph}.

By the non-renormalization theorem \cite{nonrenorm,supergraph} and
discussions in the last
section, it is sufficient for our study to
calculate the divergent parts of vertex functions
$\langle\bar{\Phi}\Phi\rangle$ and
$\langle\Phi\rangle$ with and without $\eta$ insertions.

The divergent parts of the vertex functions
$\int d^4\theta\bar{\Phi}^iT^{(\eta)j}_i\Phi_j$
and $\int d^4\theta J^{(\eta)i}\Phi_i$
in the softly broken SUSY model are
expanded in $\eta$ as
\beq
T^{(\eta)j}_i=(T+T^{(1)}\eta+T^{(1)\dagger}\bar{\eta}
+T^{(2)}\bar{\eta}\eta )_i^j, \label{eq31}
\eeq
and
\beq
J^{(\eta)i}=J^{(1)i}\bar{\eta}+J^{(2)i}\bar{\eta}\eta, \label{eq32}
\eeq
respectively.
$T_i^j$ in (\ref{eq31}) is the two-point function in the exact SUSY case.
By power counting, the factors $D\eta$ and
$\bar{D}\bar{\eta}$ do not appear in these divergent parts.
The renormalization of the coupling constants
in the lagrangian (\ref{eq23}) are then expressed as
\bea
A^{ijk}(bare)\mu^{-\epsilon}&=&A^{ijk}+\frac{1}{2}A^{i'jk}T_{i'}^i
+\frac{1}{2}A^{ij'k}T_{j'}^j+\frac{1}{2}A^{ijk'}T_{k'}^k \nonumber \\
&& -\lambda^{i'jk}T_{i'}^{(1)i}-\lambda^{ij'k}T_{j'}^{(1)j}
-\lambda^{ijk'}T_{k'}^{(1)k}+O(1/\epsilon^2), \label{eq33} \\
B^{ij}(bare)&=&B^{ij}+\frac{1}{2}B^{i'j}T_{i'}^i
+\frac{1}{2}B^{ij'}T_{j'}^j \nonumber \\
&& -M^{i'j}T_{i'}^{(1)i}-M^{ij'}T_{j'}^{(1)j}
-\lambda^{ijk}J^{(1)*}_k+O(1/\epsilon^2), \label{eq34} \\
C^{i}(bare)\mu^{\epsilon}&=&C^i+\frac{1}{2}C^{i'}T_{i'}^i
-L^{i'}T_{i'}^{(1)i}-M^{ij}J^{(1)*}_{j}+J^{(2)i}+O(1/\epsilon^2),
\label{eq35} \\
(m^2)_i^j(bare)&=&(m^2)_i^j+\frac{1}{2}(m^2)_{i'}^jT_i^{i'}
+\frac{1}{2}(m^2)_i^{j'}T_{j'}^j+T_i^{(2)j}+O(1/\epsilon^2), \label{eq36}
\eea
where $\mu$ is the renormalization scale and loop integrations are done in
$D=4-2\epsilon$ dimension. The above relations are obtained
from the relations (\ref{eq28}--\ref{eq211}) by noting
\bea
\kappa_i^j(bare)&=&-T_i^{(1)j}+O(1/\epsilon^2), \label{eq37} \\
\rho_1^i(bare)&=&-J^{(2)i}+O(1/\epsilon^2), \label{eq38} \\
\rho_2^i(bare)&=&-J^{(1)i}+O(1/\epsilon^2), \label{eq39}
\eea
in the renormalization condition $(\kappa,\rho_1,\rho_2)(renorm)=0$
in section 2.
The equations (\ref{eq33}--\ref{eq36}), which are similar to the
renormalization of the superpotential,
\beq
\lambda^{ijk}(bare)\mu^{-\epsilon}=\lambda^{ijk}
+\frac{1}{2}\lambda^{i'jk}T_{i'}^i+\frac{1}{2}\lambda^{ij'k}T_{j'}^j
+\frac{1}{2}\lambda^{ijk'}T_{k'}^k,\;\; {\rm etc.}
\eeq
are remarkable consequences of the supergraph method.

We first consider the calculation of the two-point function
$T^{(\eta)}$. This function is
obtained by inserting up to two $\eta$ operators in (\ref{eq23})
to propagators and vertices in the 1PI
supergraphs that contribute to the renormalization of
the $\int d^4\theta\bar{\Phi}\Phi$ operator.

We can commutate all $\eta$'s and $D$ operators at any step
in our calculation since the commutator $[D, \eta]=D\eta$
does not contributes to the final results (\ref{eq31}),
apart from the subdivergences for vector self-interactions
(see the last paragraph of this section).
So the $\eta$-insertions to vertices becomes almost trivial:
the $\theta$-algebra and momentum integration are completely the same
as those without the $\eta$-insertions. The insertions can be done
after all $\theta$-integrations in the supergraph have been
reduced to a single integration $\int d^4\theta$.

The $m^2\bar{\eta}{\eta}$ operator insertions to chiral supermultiplet
propagators can be treated in the same manner if we use the following
propagator with a $m^2\bar{\eta}\eta$ insertion,
\beq
\langle\Phi_i(\theta_1)\bar{\Phi}^j(\theta_2)\rangle
=\frac{i}{p^2}\delta^4(\theta_1-\theta_2)
(\delta_i^j+(m^2)_i^j\bar{\eta}\eta)+(D\eta), \label{eq310}
\eeq
instead of the exact and more involved one \cite{superfield2},
since the relevant part of (\ref{eq310}) is proportional to the
exact SUSY propagator. Note that we need not specify whether
$\eta=\theta_1^2$ or $\theta_2^2$ in (\ref{eq310}).

The gaugino mass insertions to vector supermultiplet propagators
can also be done in the same manner
after the following special care has been taken.
In exact SUSY case, we usually use the
supersymmetric gauge fixing term \cite{gaugefix,supergraph},
\beq
{\cal L}_{GF}=-\frac{1}{8\xi}\int d^4\theta(\bar{D}^2V^A)(D^2V^A),
\label{eq311}
\eeq
with the associated Fadeev-Popov ghost term,
\beq
{\cal L}_{FP}=\int d^4\theta(b+\bar{b}){\cal L}_{gV}
[(c+\bar{c})+\coth({\cal L}_{gV})(c-\bar{c})], \label{eq315}
\eeq
where
\beq
{\cal L}_{gV}X\equiv [gV, X],\;\;\;\; x\coth x=1+x^2/3-x^4/45+\cdots .
\label{eq316}
\eeq
Here $c$ and $b$ are the Fadeev-Popov ghost and anti-ghost, respectively,
which are Grassman-odd chiral superfields
in the adjoint gauge representations.
With the insertions of the gaugino mass term in (\ref{eq23}),
the propagator for vector supermultiplet is \cite{superfield2},
\bea
\langle V^A(\theta_1)V^B(\theta_2)\rangle&=&-\frac{i}{2p^2}
(1+m_A\eta+m_A^*\bar{\eta}+2|m_A|^2\bar{\eta}\eta)
\frac{D^{\alpha}\bar{D}^2D_{\alpha}}{-8p^2}\delta^4(\theta_1-\theta_2)
\delta^{AB}\nonumber \\ &&
-\frac{i\xi}{2p^2}\frac{D^2\bar{D}^2+\bar{D}^2D^2}{16p^2}
\delta^4(\theta_1-\theta_2)\delta^{AB}+(D\eta). \label{eq314}
\eea
Since the divergent vertex functions (\ref{eq31}, \ref{eq32}) are
independent of the gauge fixing parameter $\xi$, we can choose two
special values for $\xi$ in (\ref{eq314}): $\xi=0$ or $\xi=1+m_A\eta+
m_A^*\bar{\eta}+2|m_A|^2\bar{\eta}\eta$. In both cases, the propagator
becomes $(1+m_A\eta+m_A^*\bar{\eta}+2|m_A|^2\bar{\eta}\eta)$ times
the one in exact SUSY case.
So the insertion of gaugino mass term can be done as simple as
the other $\eta$-insertions.

Therefore,
the relevant supergraphs for $T^{(\eta)}$ are obtained from those
for $T$ in exact SUSY case with only dimensionless couplings,
by the following rules:
\begin{enumerate}
\item replace the $\Phi^3$ interaction vertex $\lambda^{pqr}$
by $\lambda^{pqr}-A^{pqr}\eta$,
\item replace the $\bar{\Phi}V^n\Phi(n=1,2,\cdots)$ gauge interaction
vertex $(T^n)_k^l$ by
$(T^n)_k^q(\delta_q^l-(m^2)_q^l\bar{\eta}\eta)$,
\item replace the factor $\delta_k^l$ associated with an internal
$\langle\Phi_k\bar{\Phi}^l\rangle$ line
by $\delta_k^l+(m^2)_k^l\bar{\eta}\eta$,
\item multiply an internal $\langle V^AV^A\rangle$ line
by $(1+m_A\eta+m_A^*\bar{\eta}+2|m_A|^2\bar{\eta}\eta)$, and
\item multiply the vector self-coupling $(V^A)^n(n=3,4,\cdots)$
by $(1-m_A\eta-m_A^*\bar{\eta})$.
\end{enumerate}
\beq \label{eq317} \eeq
Moreover, by a graph-theoretical argument, we can see that
$T^{(\eta)j}_i$ is obtained from the final form of $T_i^j$
by the following rules:
\begin{enumerate}
\item replace $\lambda^{pqr}$ by $\lambda^{pqr}-A^{pqr}\eta$,
\item replace gauge coupling $g_A^2$ by
$g_A^2(1+m_A\eta+m_A^*\bar{\eta}+2|m_A|^2\bar{\eta}\eta)$,
\item insert $(\delta_p^{p'}+(m^2)_p^{p'}\bar{\eta}\eta)$ between
contracted indices $p$ and $p'$ in $\lambda$ and $\lambda^*$, respectively:
$\lambda^{pqr}\lambda^*_{pq'r'}\to\lambda^{pqr}\lambda^*_{pq'r'}
+\lambda^{pqr}(m^2)_p^{p'}\lambda^*_{p'q'r'}\bar{\eta}\eta$, and
\item for terms proportional to $\delta_i^j$, multiply by
$(1-(m^2)_i^j\bar{\eta}\eta)$.
\end{enumerate}
\beq \label{eq318} \eeq

Next we consider the tadpole function $J^{(\eta)}$ in (\ref{eq32}).
By power counting and chirality conservation, the relevant supergraphs
should contain one and only one $\bar{\Phi}^2$ or
$\bar{\Phi}^2\bar{\eta}$ vertex. Therefore all the supergraphs
for $J^{(\eta)i}$ correspond one by one to those for $T^{(\eta)i}_j$
by replacing $\bar{\Phi}^2\bar{\Phi}_j$ vertex in the latter
by $\bar{\Phi}^2$. As a consequence, $J^{(\eta)i}$ is obtained from
the $\bar{\eta}$-dependent part of $T_j^{(\eta)i}$
by replacing $\lambda^*_{jkl}$ and $A^*_{jkl}$ factors in $T^{(\eta)i}_j$ by
$M^*_{kl}$ and $B^*_{kl}$, respectively.

To summarize, the divergent contributions to
two-point and tadpole $\Phi$ functions involving $\eta$
insertions, (\ref{eq31}, \ref{eq32}), are obtained from
those of the two-point functions in the exact SUSY model
by simple algebra (\ref{eq318}),
apart from subdivergences for vector self-interactions.

The renormalization group equation for the coupling constant $x_k$ is
obtained from the relation between bare and renormalized coupling
constants as
\beq
\beta(x_k)=\frac{dx_k}{d\ln\mu}+\epsilon r_kx_k=
\left(r_lx_l\frac{\partial}
{\partial x_l}-r_k \right)
a^{(1)}_k(x), \label{eq319}
\eeq
where
\beq
x_k(bare)\mu^{-r_k\epsilon}=x_k+\sum_{n=1}^{\infty}
\frac{a_k^{(n)}(x)}{\epsilon^n}, \label{eq320}
\eeq
and
\beq
r_k=\left\{ \begin{array}{cl}
1 & \;\;\mbox{\rm for}\;\;g,\;\lambda,\;A, \\
0 & \;\;\mbox{\rm for}\;\;M,\;B,\;m^2,\;m_A,\; \\
-1 & \;\;\mbox{\rm for}\;\;L,\;C. \end{array} \right. \label{eq321}
\eeq

It is easily seen that the renormalization group
equations for $A$, $B$, $C$ and $m^2$ are obtained
by multiplying the coefficients of
$\epsilon^{-1}$ on the right hand sides of (\ref{eq33}--\ref{eq36})
by 2 for the one-loop contribution, or by 4 for the two-loop contribution.

The rules (\ref{eq317}) for the $\eta$-insertion do not apply to
the vertices with only vector superfields, such as the gaugino mass term
in (\ref{eq23}) and the $\rho_3$ term in (\ref{eq213})
since these vertices contain $D$ operators in the divergent parts.
There might appear problem in the calculation of $T^{(\eta)}$
if these vector vertices appear in the supergraphs as subdivergences.
For example, in the two-loop $T^{(\eta)}$, one-loop
subdivergence of two-point vector vertex appears. Nevertheless,
we have checked
that the rules (\ref{eq317}), when applied to the one-loop $VV$ propagator,
give a correct counter term for the 1PI gaugino mass vertex,
which is zero \cite{1looplam,supergraph}.
So the rules (\ref{eq318}) are valid at least to the two-loop order.
Further study on the vector vertex functions will be reported elsewhere.

\section{One-loop renormalization group equations}
\cleqn
\pr
In this section, we calculate the one-loop $\beta$-functions
for soft SUSY breaking scalar couplings, using the supergraph method.
Our results agree with the existing results \cite{1looplam}
which have been calculated in the component field method,
but the calculation is much simpler.

The one-loop contribution to the two-point function $T_i^j$
in the exact SUSY case is \cite{1loopl}
\beq
T_i^j=\frac{1}{(4\pi)^2\epsilon}\left[
\frac{1}{2}\lambda^*_{ikl}\lambda^{jkl}-2g_A^2C_A(\Phi_i)\delta_i^j
\right], \label{eq41}
\eeq
where we adopt the notation
\beq
C_A(\Phi_i)I=R_i^AR_i^A. \label{eq42}
\eeq
Following the rules (\ref{eq318}), the divergent
vertex functions with $\eta$ are derived from (\ref{eq41}) as follows:
\bea
T_i^{(1)j}&=&\frac{1}{(4\pi)^2\epsilon}\left[
-\frac{1}{2}\lambda^*_{ikl}A^{jkl}-2g_A^2C_A(\Phi_i)m_A\delta_i^j\right],
\label{eq43}\\
T_i^{(2)j}&=&\frac{1}{(4\pi)^2\epsilon}\left[
\lambda^*_{ikl}(m^2)_{l'}^l\lambda^{jkl'}
+\frac{1}{2}A^*_{ikl}A^{jkl} \right. \nonumber \\
&& \left. -2g_A^2C_A(\Phi_i)(2|m_A|^2\delta_i^j-(m^2)_i^j)\right],
\label{eq44} \\
J^{(1)i}&=&\frac{1}{(4\pi)^2\epsilon}\left[
-\frac{1}{2}\lambda^{ikl}B^*_{kl}\right], \label{eq45} \\
J^{(2)i}&=&\frac{1}{(4\pi)^2\epsilon}\left[
(m^2)_{l'}^l\lambda^{ikl'}M^*_{kl}
+\frac{1}{2}A^{ikl}B^*_{kl}\right] . \label{eq46}
\eea
By substituting these results into (\ref{eq33}--\ref{eq36}),
we immediately obtain the $\beta$-functions
for soft SUSY breaking scalar interactions,
\bea
(4\pi)^2\beta^{(1)}(A^{ijk})&=&
\frac{1}{2}(
\lambda^{iln}\lambda^*_{i'ln}A^{i'jk}
+\lambda^{jln}\lambda^*_{j'ln}A^{ij'k}
+\lambda^{kln}\lambda^*_{k'ln}A^{ijk'}) \nonumber \\
&&+2g_A^2(C_A(\Phi_i)+C_A(\Phi_j)+C_A(\Phi_k))(2m_A\lambda^{ijk}-A^{ijk})
\nonumber \\
&&+\lambda^{i'jk}\lambda^*_{i'ln}A^{iln}
+\lambda^{ij'k}\lambda^*_{j'ln}A^{jln}
+\lambda^{ijk'}\lambda^*_{k'ln}A^{kln}, \label{eq47} \\
(4\pi)^2\beta^{(1)}(B^{ij})&=&
\frac{1}{2}(
\lambda^{iln}\lambda^*_{i'ln}B^{i'j}
+\lambda^{jln}\lambda^*_{j'ln}B^{ij'}) \nonumber \\
&&+4g_A^2C_A(\Phi_i)(2m_AM^{ij}-B^{ij})
\nonumber \\
&&+M^{i'j}\lambda^*_{i'ln}A^{iln}
+M^{ij'}\lambda^*_{j'ln}A^{jln}
+\lambda^{ijk}\lambda^*_{kln}B^{ln}, \label{eq48} \\
(4\pi)^2\beta^{(1)}(C^i)&=&
\frac{1}{2}\lambda^{iln}\lambda^*_{i'ln}C^{i'}
+L^{i'}\lambda^*_{i'ln}A^{iln}
+M^{ik}\lambda^*_{kln}B^{ln} \nonumber \\
&&+2\lambda^{ikl'}(m^2)_{l'}^lM^*_{kl}
+A^{ikl}B^*_{kl}, \label{eq49} \\
(4\pi)^2\beta^{(1)}((m^2)_i^j)&=&
\frac{1}{2}(\lambda^*_{ikl}\lambda^{i'kl}(m^2)_{i'}^j
+\lambda^*_{j'kl}\lambda^{jkl}(m^2)_i^{j'}) \nonumber \\
&&+2\lambda^*_{ikl}(m^2)_{l'}^l\lambda^{jkl'}
+A^*_{ikl}A^{jkl}
-8g_A^2C_A(\Phi_i)|m_A|^2\delta_i^j. \label{eq410}
\eea
For comparison, we also show the $\beta$-function for
$\lambda^{ijk}$,
\bea
(4\pi)^2\beta^{(1)}(\lambda^{ijk})&=&
\frac{1}{2}(
\lambda^{iln}\lambda^*_{i'ln}\lambda^{i'jk}
+\lambda^{jln}\lambda^*_{j'ln}\lambda^{ij'k}
+\lambda^{kln}\lambda^*_{k'ln}\lambda^{ijk'}) \nonumber \\
&&-2g_A^2(C_A(\Phi_i)+C_A(\Phi_j)+C_A(\Phi_k))\lambda^{ijk}. \label{eq411}
\eea

The results (\ref{eq47}--\ref{eq411}) exactly agree with the
existing results \cite{1looplam}
if the gauge group has no U(1) factor.
If the gauge group contains a U(1) factor, the following additional term
\beq
\Delta\beta^{(1)}((m^2)_i^j)=(4\pi)^{-2}2g_A^2(T^A)_i^j{\rm Tr}(T^Am^2),
\label{eq412}
\eeq
appears. This is the $\rho_3$ contribution (\ref{eq214}).

\section{Two-loop renormalization group equations}
\cleqn
\pr
In this section, we calculate the two-loop $\beta$-functions
for soft SUSY breaking scalar couplings, using the supergraph method.
We also discuss the difference between our results and
the recent results in \cite{mv5}
which are obtained in the component field method.

We should first discuss the renormalization scheme. In general,
the two-loop renormalization group equations depend on the
renormalization scheme, except for those of the gauge couplings.
In this paper, we use the \drbar scheme (dimensional reduction \cite{dr}
with modified minimal subtraction \cite{msbar}) since it is most
convenient in the supergraph calculation and respects SUSY, at least to
the two-loop order \cite{2loopdr,gooddr}.

The two-loop contribution to the two-point function
$T_i^j$ in SUSY case is
\bea
T_i^j&=&\frac{-1+\epsilon}{2(4\pi)^4\epsilon^2}\left[
4g_A^2g_B^2C_A(\Phi_i)C_B(\Phi_i)\delta_i^j
+2g_A^4C_A(\Phi_i)(T_A(\Phi)-3C_A(V))\delta_i^j \right. \nonumber \\
&&\left. +g_A^2(-C_A(\Phi_i)+2C_A(\Phi_l))\lambda^*_{ikl}\lambda^{jkl}
-\frac{1}{2}\lambda^*_{ikl}\lambda^{lst}\lambda^*_{qst}\lambda^{jkq} \right],
\label{eq51}
\eea
where we adopt the notation, in addition to (\ref{eq42}),
\bea
&& C_A(V)=C_A({\rm adj.}), \;\;
T_A(\Phi_i)\delta^{AB}= {\rm Tr}R^A_iR^B_i, \;\;
T_A(\Phi)=\sum_iT_A(\Phi_i). \label{eq52}
\eea
The result (\ref{eq51}) agrees with that in ref.\cite{2looplam}.

The divergent
contribution to $T^{(1)}$, $T^{(2)}$, $J^{(1)}$ and $J^{(2)}$ are
derived from (\ref{eq51}) by using (\ref{eq318}) as follows:
\bea
T_i^{(1)j}&=&\frac{-1+\epsilon}{2(4\pi)^4\epsilon^2}\left[
4g_A^2g_B^2C_A(\Phi_i)C_B(\Phi_i)(m_A+m_B)\delta_i^j \right.\nonumber \\
&&+4g_A^4C_A(\Phi_i)(T_A(\Phi)-3C_A(V))m_A\delta_i^j \nonumber \\
&&+g_A^2(-C_A(\Phi_i)+2C_A(\Phi_l))(\lambda^*_{ikl}\lambda^{jkl}m_A
-\lambda^*_{ikl}A^{jkl}) \nonumber \\
&&\left. +\frac{1}{2}(
\lambda^*_{ikl}A^{lst}\lambda^*_{qst}\lambda^{jkq}
+\lambda^*_{ikl}\lambda^{lst}\lambda^*_{qst}A^{jkq})\right], \label{eq53} \\
T_i^{(2)j}&=&\frac{-1+\epsilon}{2(4\pi)^4\epsilon^2}\left[
4g_A^2g_B^2C_A(\Phi_i)C_B(\Phi_i)((2|m_A|^2+2|m_B|^2+m_Am_B^*+m_A^*m_B)
\delta_i^j-(m^2)_i^j) \right.\nonumber \\
&&+2g_A^4C_A(\Phi_i)(T_A(\Phi)-3C_A(V))
(6|m_A|^2\delta_i^j-(m^2)_i^j) \nonumber \\
&&+g_A^2(-C_A(\Phi_i)+2C_A(\Phi_l))(
2\lambda^*_{ikl}\lambda^{jkl}|m_A|^2
-A^*_{ikl}\lambda^{jkl}m_A
-\lambda^*_{ikl}A^{jkl}m_A^* \nonumber \\
&&+A^*_{ikl}A^{jkl}
+\lambda^*_{ikl}(m^2)_{k'}^k\lambda^{jk'l}
+\lambda^*_{ikl}(m^2)_{l'}^l\lambda^{jkl'}) \nonumber \\
&&-\frac{1}{2}(
A^*_{ikl}A^{lst}\lambda^*_{qst}\lambda^{jkq}
+A^*_{ikl}\lambda^{lst}\lambda^*_{qst}A^{jkq}
+\lambda^*_{ikl}A^{lst}A^*_{qst}\lambda^{jkq} \nonumber \\
&& +\lambda^*_{ikl}\lambda^{lst}A^*_{qst}A^{jkq}
+\lambda^*_{ikl}(m^2)_{k'}^k\lambda^{lst}\lambda^*_{qst}\lambda^{jk'q}
+\lambda^*_{ikl}(m^2)_{l'}^l\lambda^{l'st}\lambda^*_{qst}\lambda^{jkq}
\nonumber \\
&&\left.
+\lambda^*_{ikl}\lambda^{lst}\lambda^*_{qst}(m^2)_{q'}^q\lambda^{jkq'}
+2\lambda^*_{ikl}\lambda^{lst}(m^2)_t^{t'}\lambda^*_{qst'}\lambda^{jkq}
)\right], \label{eq54}
\eea
and
\bea
J^{(1)i}&=&\frac{-1+\epsilon}{2(4\pi)^4\epsilon^2}\left[
2g_A^2C_A(\Phi_l)(\lambda^{ikl}M^*_{kl}m_A^*
-\lambda^{ikl}B^*_{kl}) \right. \nonumber \\
&&\left. +\frac{1}{2}(
\lambda^{ikq}A^*_{qst}\lambda^{lst}M^*_{kl}
+\lambda^{ikq}\lambda^*_{qst}\lambda^{lst}B^*_{kl})\right], \label{eq55} \\
J^{(2)i}&=&\frac{-1+\epsilon}{2(4\pi)^4\epsilon^2}\left[
2g_A^2C_A(\Phi_l)(
2\lambda^{ikl}M^*_{kl}|m_A|^2
-\lambda^{ikl}B^*_{kl}m_A
-A^{ikl}M^*_{kl}m_A^* \right.\nonumber \\
&&+A^{ikl}B^*_{kl}
+\lambda^{ik'l}(m^2)_{k'}^kM^*_{kl}
+\lambda^{ikl'}(m^2)_{l'}^lM^*_{kl}) \nonumber \\
&&-\frac{1}{2}(
\lambda^{ikq}\lambda^*_{qst}A^{lst}B^*_{kl}
+A^{ikq}\lambda^*_{qst}\lambda^{lst}B^*_{kl}
+\lambda^{ikq}A^*_{qst}A^{lst}M^*_{kl} \nonumber \\
&& +A^{ikq}A^*_{qst}\lambda^{lst}M^*_{kl}
+\lambda^{ik'q}(m^2)_{k'}^k\lambda^*_{qst}\lambda^{lst}M^*_{kl}
+\lambda^{ikq}\lambda^*_{qst}\lambda^{l'st}(m^2)_{l'}^lM^*_{kl}
\nonumber \\
&&\left. +\lambda^{ikq'}(m^2)_{q'}^q\lambda^*_{qst}\lambda^{lst}M^*_{kl}
+2\lambda^{ikq}\lambda^*_{qst'}(m^2)_t^{t'}\lambda^{lst}M^*_{kl}
)\right]. \label{eq56}
\eea

By substituting these results into (\ref{eq33}--\ref{eq36}),
we obtain the two-loop $\beta$-functions for soft SUSY
breaking scalar interactions in the
\drbar scheme.
The results are listed below:
\bea
(4\pi)^4\beta^{(2)}(A^{ijk})&=&
4g_A^2g_B^2(C_A(\Phi_i)C_B(\Phi_i)+C_A(\Phi_j)C_B(\Phi_j)
+C_A(\Phi_k)C_B(\Phi_k)) \nonumber \\
&&\times (A^{ijk}-2(m_A+m_B)\lambda^{ijk}) \nonumber \\
&&+2g_A^4(C_A(\Phi_i)+C_A(\Phi_j)+C_A(\Phi_k))
(T_A(\Phi)-3C_A(V)) \nonumber \\
&&\times (A^{ijk}-4m_A\lambda^{ijk}) \nonumber \\
&&+g_A^2
(-C_A(\Phi_i)+2C_A(\Phi_l))\lambda^{iln}\lambda^*_{i'ln}A^{i'jk} \nonumber \\
&&+g_A^2
(-C_A(\Phi_j)+2C_A(\Phi_l))\lambda^{jln}\lambda^*_{j'ln}A^{ij'k} \nonumber \\
&&+g_A^2
(-C_A(\Phi_k)+2C_A(\Phi_l))\lambda^{kln}\lambda^*_{k'ln}A^{ijk'} \nonumber \\
&&-\frac{1}{2}(
\lambda^{inq}\lambda^*_{qst}\lambda^{lst}\lambda^*_{i'nl}A^{i'jk}
+\lambda^{jnq}\lambda^*_{qst}\lambda^{lst}\lambda^*_{j'nl}A^{ij'k}
+\lambda^{knq}\lambda^*_{qst}\lambda^{lst}\lambda^*_{k'nl}A^{ijk'}
) \nonumber \\
&&+2g_A^2(\lambda^{inl}m_A-A^{inl})(C_A(\Phi_i)-2C_A(\Phi_l))
\lambda^*_{i'nl}\lambda^{i'jk} \nonumber \\
&&+2g_A^2(\lambda^{jnl}m_A-A^{jnl})(C_A(\Phi_j)-2C_A(\Phi_l))
\lambda^*_{j'nl}\lambda^{ij'k} \nonumber \\
&&+2g_A^2(\lambda^{knl}m_A-A^{knl})(C_A(\Phi_k)-2C_A(\Phi_l))
\lambda^*_{k'nl}\lambda^{ijk'} \nonumber \\
&&-(\lambda^{inq}\lambda^*_{qst}A^{lst}\lambda^*_{i'nl}
+A^{inq}\lambda^*_{qst}\lambda^{lst}\lambda^*_{i'nl})\lambda^{i'jk}
\nonumber \\
&&-(\lambda^{jnq}\lambda^*_{qst}A^{lst}\lambda^*_{j'nl}
+A^{jnq}\lambda^*_{qst}\lambda^{lst}\lambda^*_{j'nl})\lambda^{ij'k}
\nonumber \\
&&-(\lambda^{knq}\lambda^*_{qst}A^{lst}\lambda^*_{k'nl}
+A^{knq}\lambda^*_{qst}\lambda^{lst}\lambda^*_{k'nl})\lambda^{ijk'},
\label{eq57} \\
(4\pi)^4\beta^{(2)}(B^{ij})&=&
8g_A^2g_B^2C_A(\Phi_i)C_B(\Phi_i)(B^{ij}-2(m_A+m_B)M^{ij}) \nonumber \\
&&+4g_A^4C_A(\Phi_i)(T_A(\Phi)-3C_A(V))(B^{ij}-4m_AM^{ij}) \nonumber \\
&&+g_A^2(-C_A(\Phi_i)+2C_A(\Phi_l))\lambda^{ikl}\lambda^*_{i'kl}B^{i'j}
\nonumber \\
&&+g_A^2(-C_A(\Phi_j)+2C_A(\Phi_l))\lambda^{jkl}\lambda^*_{j'kl}B^{ij'}
\nonumber \\
&&-\frac{1}{2}(
\lambda^{inq}\lambda^*_{qst}\lambda^{lst}\lambda^*_{i'nl}B^{i'j}
+\lambda^{jnq}\lambda^*_{qst}\lambda^{lst}\lambda^*_{j'nl}B^{ij'})
\nonumber \\
&&+2g_A^2(\lambda^{inl}m_A-A^{inl})(C_A(\Phi_i)-2C_A(\Phi_l))
\lambda^*_{i'nl}M^{i'j} \nonumber \\
&&+2g_A^2(\lambda^{jnl}m_A-A^{jnl})(C_A(\Phi_j)-2C_A(\Phi_l))
\lambda^*_{j'nl}M^{ij'} \nonumber \\
&&-(
\lambda^{inq}\lambda^*_{qst}A^{lst}\lambda^*_{i'nl}
+A^{inq}\lambda^*_{qst}\lambda^{lst}\lambda^*_{i'nl})M^{i'j} \nonumber \\
&&-(
\lambda^{jnq}\lambda^*_{qst}A^{lst}\lambda^*_{j'nl}
+A^{jnq}\lambda^*_{qst}\lambda^{lst}\lambda^*_{j'nl})M^{ij'} \nonumber \\
&&-4g_A^2C_A(\Phi_l)\lambda^*_{knl}
(M^{nl}m_A-B^{nl})\lambda^{ijk} \nonumber \\
&&-(\lambda^*_{knq}A^{qst}\lambda^*_{lst}M^{nl}
+\lambda^*_{knq}\lambda^{qst}\lambda^*_{lst}B^{nl})\lambda^{ijk},
\label{eq58} \\
(4\pi)^4\beta^{(2)}(C^i)&=&
2g_A^2C_A(\Phi_l)\lambda^{ikl}\lambda^*_{i'kl}C^{i'}
-\frac{1}{2}\lambda^{ikq}\lambda^*_{qst}\lambda^{lst}\lambda^*_{i'kl}C^{i'}
\nonumber \\
&&-4g_A^2C_A(\Phi_l)(\lambda^{ikl}m_A-A^{ikl})
\lambda^*_{i'kl}L^{i'} \nonumber \\
&&-(\lambda^{ikq}\lambda^*_{qst}A^{lst}\lambda^*_{i'kl}
+A^{ikq}\lambda^*_{qst}\lambda^{lst}\lambda^*_{i'kl})L^{i'}
\nonumber \\
&&-4g_A^2C_A(\Phi_l)\lambda^*_{jnl}
(M^{nl}m_A-B^{nl})M^{ij} \nonumber \\
&&-(\lambda^*_{jnq}A^{qst}\lambda^*_{lst}M^{nl}
+\lambda^*_{jnq}\lambda^{qst}\lambda^*_{lst}B^{nl})M^{ij}  \nonumber \\
&&+4g_A^2C_A(\Phi_l)(
2\lambda^{ikl}M^*_{kl}|m_A|^2
-\lambda^{ikl}B^*_{kl}m_A
-A^{ikl}M^*_{kl}m_A^* \nonumber \\
&&+A^{ikl}B^*_{kl}
+\lambda^{ik'l}(m^2)_{k'}^kM^*_{kl}
+\lambda^{ikl'}(m^2)_{l'}^lM^*_{kl}) \nonumber \\
&&-(
\lambda^{ikq}\lambda^*_{qst}A^{lst}B^*_{kl}
+A^{ikq}\lambda^*_{qst}\lambda^{lst}B^*_{kl}
+\lambda^{ikq}A^*_{qst}A^{lst}M^*_{kl} \nonumber \\
&& +A^{ikq}A^*_{qst}\lambda^{lst}M^*_{kl}
+\lambda^{ik'q}(m^2)_{k'}^k\lambda^*_{qst}\lambda^{lst}M^*_{kl}
+\lambda^{ikq}\lambda^*_{qst}\lambda^{l'st}(m^2)_{l'}^lM^*_{kl}
\nonumber \\
&& +\lambda^{ikq'}(m^2)_{q'}^q\lambda^*_{qst}\lambda^{lst}M^*_{kl}
+2\lambda^{ikq}\lambda^*_{qst'}(m^2)_t^{t'}\lambda^{lst}M^*_{kl} ),
\label{eq59} \\
(4\pi)^4\beta^{(2)}((m^2)_i^j)&=&
8g_A^2g_B^2C_A(\Phi_i)C_B(\Phi_i)(2|m_A|^2+2|m_B|^2+m_Am_B^*+m_A^*m_B)
\delta_i^j \nonumber \\
&&+24g_A^4C_A(\Phi_i)(T_A(\Phi)-3C_A(V))
|m_A|^2\delta_i^j \nonumber \\
&&+(-g_A^2C_A(\Phi_i)+2g_A^2C_A(\Phi_l))
(\lambda^*_{ikl}\lambda^{i'kl}(m^2)_{i'}^j
+\lambda^*_{j'kl}\lambda^{jkl}(m^2)_i^{j'}) \nonumber \\
&&-\frac{1}{2}(
\lambda^*_{ikl}\lambda^{lst}\lambda^*_{qst}\lambda^{i'kq}(m^2)_{i'}^j
+\lambda^*_{j'kl}\lambda^{lst}\lambda^*_{qst}\lambda^{jkq}(m^2)_i^{j'} )
\nonumber \\
&& +2g_A^2(-C_A(\Phi_i)+2C_A(\Phi_l))(
2\lambda^*_{ikl}\lambda^{jkl}|m_A|^2
-A^*_{ikl}\lambda^{jkl}m_A
-\lambda^*_{ikl}A^{jkl}m_A^* \nonumber \\
&&+A^*_{ikl}A^{jkl}
+\lambda^*_{ikl}(m^2)_{k'}^k\lambda^{jk'l}
+\lambda^*_{ikl}(m^2)_{l'}^l\lambda^{jkl'}) \nonumber \\
&&-
A^*_{ikl}A^{lst}\lambda^*_{qst}\lambda^{jkq}
-A^*_{ikl}\lambda^{lst}\lambda^*_{qst}A^{jkq}
-\lambda^*_{ikl}A^{lst}A^*_{qst}\lambda^{jkq} \nonumber \\
&& -\lambda^*_{ikl}\lambda^{lst}A^*_{qst}A^{jkq}
-\lambda^*_{ikl}(m^2)_{k'}^k\lambda^{lst}\lambda^*_{qst}\lambda^{jk'q}
-\lambda^*_{ikl}(m^2)_{l'}^l\lambda^{l'st}\lambda^*_{qst}\lambda^{jkq}
\nonumber \\
&& -\lambda^*_{ikl}\lambda^{lst}\lambda^*_{qst}(m^2)_{q'}^q\lambda^{jkq'}
-2\lambda^*_{ikl}\lambda^{lst}(m^2)_t^{t'}\lambda^*_{qst'}\lambda^{jkq}.
\label{eq510}
\eea
For comparison, we show the $\beta$-function for
$\lambda^{ijk}$,
\bea
(4\pi)^4\beta^{(2)}(\lambda^{ijk})&=&
4g_A^2g_B^2(C_A(\Phi_i)C_B(\Phi_i)+C_A(\Phi_j)C_B(\Phi_j)
+C_A(\Phi_k)C_B(\Phi_k))\lambda^{ijk} \nonumber \\
&&+2g_A^4(C_A(\Phi_i)+C_A(\Phi_j)+C_A(\Phi_k))
(T_A(\Phi)-3C_A(V))\lambda^{ijk} \nonumber \\
&&+g_A^2(-C_A(\Phi_i)+2C_A(\Phi_l))
\lambda^{inl}\lambda^*_{i'nl}\lambda^{i'jk} \nonumber \\
&&+g_A^2(-C_A(\Phi_j)+2C_A(\Phi_l))
\lambda^{jnl}\lambda^*_{j'nl}\lambda^{ij'k} \nonumber \\
&&+g_A^2(-C_A(\Phi_k)+2C_A(\Phi_l))
\lambda^{knl}\lambda^*_{k'nl}\lambda^{ijk'} \nonumber \\
&&-\frac{1}{2}(
\lambda^{inq}\lambda^*_{qst}\lambda^{lst}\lambda^*_{i'nl}\lambda^{i'jk}
+\lambda^{jnq}\lambda^*_{qst}\lambda^{lst}\lambda^*_{j'nl}\lambda^{ij'k}
\nonumber \\
&&+\lambda^{knq}\lambda^*_{qst}\lambda^{lst}\lambda^*_{k'nl}\lambda^{ijk'}
 ). \label{eq511}
\eea
As in the last section, the contribution of $\rho_3$ (\ref{eq214}) is not
included in $\beta(m^2)$.

Finally, we compare our results to the other calculation. Recently,
the two-loop $\beta$-functions for $A$, $B$ and $m^2$ have been obtained in
ref.\cite{mv5} by using the component field method.
They have used the following method: apply the general
formulas given in \cite{mv} to the softly broken SUSY model
in the Wess-Zumino gauge and convert
them to the \drbar scheme by the one-loop finite
transformation \cite{2loopmg2}.
Our results agree with theirs for $A$ and $B$, but
do not for $m^2$. The difference is
\bea
&&(4\pi)^4\beta^{(2)}((m^2)_i^j)[6]-{\rm eq.}(5.10)= \nonumber \\
&&-2g_A^2(T^A)_i^j(T^A)_p^q(m^2)_q^r\lambda^*_{rst}\lambda^{pst}
+8g_A^2g_B^2(T^A)_i^j{\rm Tr}(T^AC_B(\Phi_r)m^2) \nonumber \\
&&+8g_A^4C_A(\Phi_i){\rm Tr}(T_A(\Phi_r)m^2)\delta_i^j
-8g_A^4C_A(\Phi_i)C_A(V)|m_A|^2\delta_i^j .
\eea
The first two terms are the contributions from $\rho_3$ (\ref{eq214})
that we neglected in the present analysis and they cause no problem.
The remaining two terms show, however, a serious discrepancy between
our results and those of ref.\cite{mv5}.
The discrepancy can be clearly shown for the
following simple case: $G$ is simple, $\lambda=M=L=0$, $A=B=C=m_A=0$ and
$(m^2)_i^j=m_i^2\delta_i^j$. The renormalization group equation for $m^2$
is then
\beq
(4\pi)^4\frac{dm_i^2}{d\ln \mu}=\left\{
\begin{array}{l} 0,\;\;\;(5.10) \\
8g_A^4C_A(\Phi_i)\sum_r T_A(\Phi_r)m_r^2.\;\;\;({\rm ref.}[6])
\end{array} \right.
\eeq
We have found by a direct calculation in the \drbar scheme
that, for this simple case, the component field method does
reproduce our result. However, we have so far been unable to
trace the origin of the discrepancy and the definite conclusion
is left to further studies.

\section{Conclusion}
\pr
In this paper, we have obtained the renormalization
group equations for soft SUSY breaking scalar couplings
to the two-loop order, by using the supergraph method.
We have shown that, due to the non-renormalization theorem,
the renormalization group equations for these
couplings are obtained
by evaluating one- and two-point functions of chiral superfields
involving the spurion field $\eta$,
if the gauge group has no U(1) factor.
Under the same condition, we have also shown that
the one- and two-loop relevant divergent terms involving
$\eta$ can be obtained  by simple algebra from those
in exact SUSY case. Our results basically agree with the existing results
which are obtained by the component field method, but the calculation
is much simpler once the SUSY two-point function $T_i^j$ in (\ref{eq31})
has been given.
A discrepancy in the two-loop $\beta$-function for the scalar mass term is
found between our results and those of ref.\cite{mv5} that are obtained
by finite transformation from the non-SUSY $\beta$-functions of the
component field.
Although we confirmed in a simple case the validity of our result by an
explicit calculation in the component field method, we have not been able
to trace the origin of the discrepancy.

Our method of obtaining the $\beta$-functions for the soft SUSY
breaking terms from the exact SUSY results might be extended to
higher loop orders, if the following problems are solved:
the potential inconsistency \cite{baddr} of the \drbar scheme
and the renormalization of the vector and ghost superfields in softly
broken SUSY models.

If the gauge group has a U(1) factor,
such as the standard model, we should add
the $\rho_3$ contribution (\ref{eq214}) to $\beta(m^2)$.
The analysis including this contribution will be reported elsewhere.

\section*{Note added in proof}
\pr
After this paper was accepted for publication, we received a paper
[21] which found a different result for $\beta^{(2)}(m^2)$. Ref.[21]
points out that for the proper renormalization in the \drbar scheme,
the mass terms $\tilde{m}_A$ for the $\epsilon$-scalars
(the last $2\epsilon$ components of the vector
fields $V^A_{\mu}$) and their counterterms should be included into
${\cal L}_{soft}$. As a result, Eq.(5.10) is modified as
\begin{eqnarray*}
(4\pi)^4\beta^{(2)}_{\drbar}((m^2)_i^j)&=&(5.10)+
(\rho_3-\mbox{contribution}) \nonumber\\
&&+8\delta_i^jg_A^4C_A(\Phi_i)\left[ 2{\rm Tr}(C_A(\Phi_r)m^2)-2C_A(V)|m_A|^2
+(T_A(\Phi)-3C_A(V))\tilde{m}_A^2\right] \nonumber\\
&&-2g_A^2(C_A(\Phi_i)+2C_A(\Phi_l))\lambda^*_{ikl}\lambda^{jkl}\tilde{m}_A^2
\end{eqnarray*}
We have checked that this result is obtained in the superfiled
formalism after inclusion of the $\epsilon$-scalar masses,
\[
\frac{\tilde{m}_A^2}{2}V^A_{\mu}V^A_{\nu}\hat{\hat{g}}^{\mu\nu}=
\frac{\tilde{m}_A^2}{2}\int d^4\theta\bar{\eta}\eta
\frac{1}{16g^2}
\bar{\sigma}_{\mu}^{\dot{\alpha}\alpha}\bar{D}_{\dot{\alpha}}
(e^{-2gV}D_{\alpha}e^{2gV})
\bar{\sigma}_{\nu}^{\dot{\beta}\beta}\bar{D}_{\dot{\beta}}
(e^{-2gV}D_{\beta}e^{2gV})
\hat{\hat{g}}^{\mu\nu}.
\]
Here $\hat{\hat{g}}^{\mu\nu}$ is the $2\epsilon$-dimensional metric tensor.
In the superfield formalism,
the extra contribution to $\beta^{(2)}_{\drbar}(m^2)$ above
come from the additional terms to
$T^{(2)}$ in (3.1), which are produced by the $\tilde{m}^2$ insertion
to supergraphs and
by the subdivergence $\bar{\eta}\eta(\bar{D}\bar{\sigma}DV)^2$
which is absent in naive calculation.
We have found that the corresponding additional terms to $J^{(2)}$
in (3.2) modifies $\beta^{(2)}(C)$ of (5.9) as,
\[
(4\pi)^4\beta^{(2)}_{\drbar}(C^i)=(5.9)
-4g_A^2C_A(l)\lambda^{ikl}M^*_{kl}\tilde{m}_A^2.
\]
Details of the discussion will be presented elsewhere.

We thank I. Jack, D.R.T. Jones, S.P. Martin and M.T. Vaughn for
clarifying discussions on this problem.

\section*{\large Acknowledgements}

We would like to thank K.~Fujikawa, K.~Hagiwara, K.~Hikasa, K.~Inoue,
H.~Kawai, H.~Murayama and Y.~Okada
for fruitful discussions. We also thank Soryushi Shogakukai for
financial support.

\section*{Appendix}
\renewcommand{\theequation}{A.\arabic{equation}}
\cleqn
\pr
Here we list our notations and conventions for superfield formalism.

The conventions for 2-component spinors are
\[ \epsilon^{\alpha\beta}=\pm 1 =-\epsilon_{\alpha\beta},\;\;\;
\epsilon^{\dot{\alpha}\dot{\beta}}=(\epsilon^{\alpha\beta})^*, \]
\[ \psi^{\alpha}=\epsilon^{\alpha\beta}\psi_{\beta},\;\;\;
\psi_{\alpha}=\epsilon_{\alpha\beta}\psi^{\beta}, \]
\[ \psi\xi=\psi^{\alpha}\xi_{\alpha},\;\;\;
\bar{\psi}\bar{\xi}=\bar{\psi}_{\alpha}\bar{\xi}^{\alpha}, \]
\[ \sigma^{\mu}_{\alpha\dot{\alpha}}=(1, \sigma_{1-3}),\;\;\;
\bar{\sigma}^{\mu\dot{\alpha}\alpha}=
\epsilon^{\alpha\beta}\epsilon^{\dot{\alpha}\dot{\beta}}
\sigma^{\mu}_{\beta\dot{\beta}}, \]
\beq g_{\mu\nu}=(+,-,-,-). \eeq
The notations for superfields are
\[ \int d^2\theta\;\theta^2=1,\;\;\;
\int d^4\theta\;\bar{\theta}^2\theta^2=1,\]
\[ D_{\alpha}=\frac{\partial}{\partial\theta^{\alpha}}
-i(\sigma^{\mu}\bar{\theta})_{\alpha}\partial_{\mu},\;\;\;
\bar{D}_{\dot{\alpha}}=-\frac{\partial}{\partial\bar{\theta}^{\dot{\alpha}}}
+i(\theta\sigma^{\mu})_{\dot{\alpha}}\partial_{\mu},\]
\beq \{ D_{\alpha}, \bar{D}_{\dot{\alpha}} \} =
2i\sigma^{\mu}_{\alpha\dot{\alpha}}\partial_{\mu}. \eeq
By using the following decompositions of the superfields
in the Wess-Zumino gauge,
\bea
\Phi(\theta)&=&\phi+\sqrt{2}\theta\psi+\theta^2F, \nonumber \\
V(\theta)&=&-\theta\sigma^{\mu}\bar{\theta}V_{\mu}
+\bar{\theta}^2\theta\chi+\theta^2\bar{\theta}\bar{\chi}
+\frac{1}{2}\theta^2\bar{\theta}^2D',
\eea
we can show that eq.(\ref{eq21}) gives the properly normalized
lagrangian,
\bea
{\cal L}_{SUSY}&=& -\frac{1}{4}
(\partial_{\mu}V_{\nu}-\partial_{\nu}V_{\mu}-ig[V_{\mu}, V_{\nu}])^2
+i\chi\sigma^{\mu}(\partial_{\mu}\bar{\chi}-ig[V_{\mu}, \bar{\chi}])
+\frac{1}{2}D'^2 \nonumber \\
&&+|(\partial_{\mu}-igV_{\mu})\phi|^2+i\bar{\psi}\bar{\sigma}^{\mu}
(\partial_{\mu}-igV_{\mu})\psi+|F|^2 \nonumber \\
&&-\sqrt{2}g(\phi^{\dagger}\chi\psi+\bar{\psi}\bar{\chi}\phi)
+g\phi^{\dagger}D'\phi \nonumber \\
&&+\left( \frac{\partial W(\phi)}{\partial\phi_i}F_i
-\frac{1}{2}\frac{\partial^2 W(\phi)}{\partial\phi_i\partial\phi_j}
\psi_i\psi_j\right)+{\rm h.c.},
\eea
where
\beq
W(\phi)=\frac{1}{6}\lambda^{ijk}\phi_i\phi_j\phi_k
+\frac{1}{2}M^{ij}\phi_i\phi_j+L^i\phi_i,
\eeq
and
\bea
{\cal L}_{soft}&=&-\left( \frac{1}{6}A^{ijk}\phi_i\phi_j\phi_k+
\frac{1}{2}B^{ij}\phi_i\phi_j+C^i\phi_i+\frac{m_A}{2}\chi^A\chi^A\right)
+{\rm h.c.}\nonumber \\
&&-\phi^{*i}(m^2)_i^j\phi_j.
\eea


\def\PL #1 #2 #3 {Phys.~Lett. {\bf#1}, #2 (#3) }
\def\NP #1 #2 #3 {Nucl.~Phys. {\bf#1}, #2 (#3) }
\def\ZP #1 #2 #3 {Z.~Phys. {\bf#1}, #2 (#3) }
\def\PR #1 #2 #3 {Phys.~Rev. {\bf#1}, #2 (#3) }
\def\PP #1 #2 #3 {Phys.~Rep. {\bf#1}, #2 (#3) }
\def\PRL #1 #2 #3 {Phys.~Rev.~Lett. {\bf#1}, #2 (#3) }
\def\PTP #1 #2 #3 {Prog.~Theor.~Phys. {\bf#1}, #2 (#3) }
\def\ib #1 #2 #3 {{\it ibid.} {\bf#1}, #2 (#3) }
\def\etal {{\it et al}.}
\def\eg {{\it e.g}.}
\def\ie {{\it i.e}.}


\newpage
\section*{\large \bf References}
\begin{thebibliography}{99}

\bibitem{1looplam}
K. Inoue, A. Kakuto, H. Komatsu and S. Takeshita, \PTP 68 927 1982
(erratum \ib 70 330 1983 ) ; \ib 71 413 1984 ;\\
J.P. Derendinger and C.A. Savoy, \NP B237 307 1984 ;\\
B. Gato, J. Le\'on, J. P\'erez-Mercader and M. Quir\'os,
\NP B253 285 1985 ;\\
N.K. Falck, \ZP C30 247 1986 .

\bibitem{mv}
M.E. Machacek and M.T. Vaughn, \NP B222 83 1983 ; \ib B236 221 1984 ;
\ib B249 70 1985 .

\bibitem{2loopdr}
I. Jack, \PL B147 405 1984 .

\bibitem{2loopmg1}
Y. Yamada, \PL B316 109 1993 ; \PRL 72 25 1994 .

\bibitem{2loopmg2}
S.P. Martin and M.T. Vaughn, \PL B318 331 1993 .

\bibitem{mv5}
S.P. Martin and M.T. Vaughn, Northeastern University preprint
NUB-3081-93TH (to be published in Phys. Rev. D) .

\bibitem{supergraph1}
R. Delbourgo, Nuovo Cimento {\bf 25A}, 646 (1975);\\
A. Salam and J. Strathdee, \NP B86 142 1975 ;\\
K. Fujikawa and W. Lang, \NP B88 61 1975 .

\bibitem{supergraph}
M.T. Grisaru, M. Ro\v{c}ek and W. Siegel, \NP B159 429 1979 .

\bibitem{soft}
L. Girardello and M.T. Grisaru, \NP B194 65 1982 .

\bibitem{superfield2}
J.A. Helay\"el-Neto, \PL B135 78 1984 ;\\
F. Feruglio, J.A. Helay\"el-Neto and F. Legovini,\\
\NP B249 533 1985 ;\\
M. Scholl, \ZP C28 545 1985 .

\bibitem{nonrenorm}
J. Wess and B. Zumino, \PL B49 52 1974 ;\\
I. Iliopoulos and B. Zumino, \NP B76 310 1974 .

\bibitem{chang}
D. Chang and A. Gangopadhyaya, \PR D33 1771 1986 .

\bibitem{superfield1}
S. Ferrara and B. Zumino, \NP B79 413 1974 ;\\
A. Salam and J. Strathdee, \PL B51 353 1974 ; \PR D11 1521 1974 .

\bibitem{gaugefix}
S. Ferrara and O. Piguet, \NP B93 261 1975 .

\bibitem{1loopl}
R. Barbieri, S. Ferrara, L. Maiani, F. Palambo and C.A. Savoy,
\PL B115 212 1982 .

\bibitem{dr}
W. Siegel, \PL B84 193 1979 ;\\
D.M. Capper, D.R.T. Jones and P. van Nieuwenhuizen, \NP B167 479 1980 .

\bibitem{msbar}
W.A. Bardeen, A.J. Buras, D.W. Duke and T. Muta,
\PR D18 3998 1978 .

\bibitem{gooddr}
P. Howe, A. Parkes and P. West, \PL B147 409 1984 .

\bibitem{2looplam}
P. West, \PL B137 371 1984 ;\\
D.R.T. Jones and L. Mezincescu, \PL B138 293 1984 .

\bibitem{baddr}
W. Siegel, \PL B94 37 1980 ;\\
L.V. Avdeev and A.A. Vladimirov, \NP B219 262 1983 .

\bibitem{ref21}
I. Jack and D.R.T. Jones, Liverpool University preprint
LTH-334 (1994) .

\end{thebibliography}
\end{document}